\documentclass[printer]{aa}
\usepackage[utf8]{inputenc}
\usepackage{natbib}
\usepackage{graphicx}
\usepackage{amsmath}
\usepackage{gensymb}
\usepackage{subcaption}
\usepackage{dblfloatfix}
\usepackage{makecell}
\usepackage{ragged2e}
\usepackage[colorlinks=true,allcolors=blue]{hyperref}
\usepackage[scientific-notation=true]{siunitx}
\usepackage{xcolor}

\title{A dichotomy in group II Herbig disks}
\subtitle{ALMA gas disk height measurements show both shadowed large vertically extended disks and compact flat disks}

\author{L. M. Stapper\inst{\ref{inst1}} \and M. R. Hogerheijde\inst{\ref{inst1}, \ref{inst2}} \and E. F. van Dishoeck\inst{\ref{inst1},\ref{inst3}} \and T. Paneque-Carreño\inst{\ref{inst1}, \ref{inst4}}}

\institute{Leiden Observatory, Leiden University, PO Box 9513, 2300 RA Leiden, The Netherlands \\e-mail:\texttt{stapper@strw.leidenuniv.nl} \label{inst1} \and Anton Pannekoek Institute for Astronomy, University of Amsterdam, PO Box 94249, 1090 GE, Amsterdam, The Netherlands \label{inst2} \and Max-Planck-Institut für Extraterrestrische Physik, Giessenbachstrasse 1, 85748 Garching, Germany \label{inst3} \and European Southern Observatory, Karl-Schwarzschild-Str 2, 85748 Garching, Germany \label{inst4}}

\date{\today}

\abstract
{ %Context
Herbig stars can be classified into group I and group II depending on the shape of the far-IR excess from the spectral energy distribution. This separation may be evolutionary and related to the vertical structure of these disks.
}
{ %Aims
We aim to determine the emission height of Herbig disks and compare the resulting vertical extent of both groups.
}
{ %Methods
ALMA Band 6 observations of $^{12}$CO $J$=2-1 emission lines at sufficient velocity ($\sim0.3$~km~s$^{-1}$) and spatial resolution ($\sim30$~au) of eight Herbig disks (four group I and four group II sources) are used to determine the emission heights from the channel maps via geometrical methods previously developed in other works.
}
{ %Results
We find that all group I disks are vertically extended with a height to radius ratio of at least 0.25, and for three of the disks the gas emission profile can be traced out to 200-500~au. The group II disks are divided between MWC~480 and HD~163296 which have similar emission height profiles as the group I disks and AK~Sco and HD~142666 which are very flat (not exceeding a height of 10~au over the full extent traced) and more compact ($<200$~au in size). The brightness temperatures show no differences between the disks when the luminosity of the host star is accounted for.
}
{ %Conclusions
Our findings agree with previous work suggesting that group I disks are vertically extended and that group II disks are either large and self-shadowed or compact. Both MWC~480 and HD~163296 could be precursors of group I disks, which we see now before a cavity has formed that would allow irradiation of the outer parts of the disk. The very flat disks AK~Sco and HD~142666 could be due to significant settling because of the advanced age of these disks ($\sim20$ instead of $<10$~Myr). These large differences in vertical structures are not reflected in the spectral energy distributions of these disks. More and deeper observations at higher spatial and velocity resolution are necessary to further characterize the Herbig sub-groups.
}


\begin{document}

\maketitle

\section{Introduction}
\label{sec:introduction}

\begin{figure*}[t]
    \centering
    \includegraphics[width=0.8\textwidth]{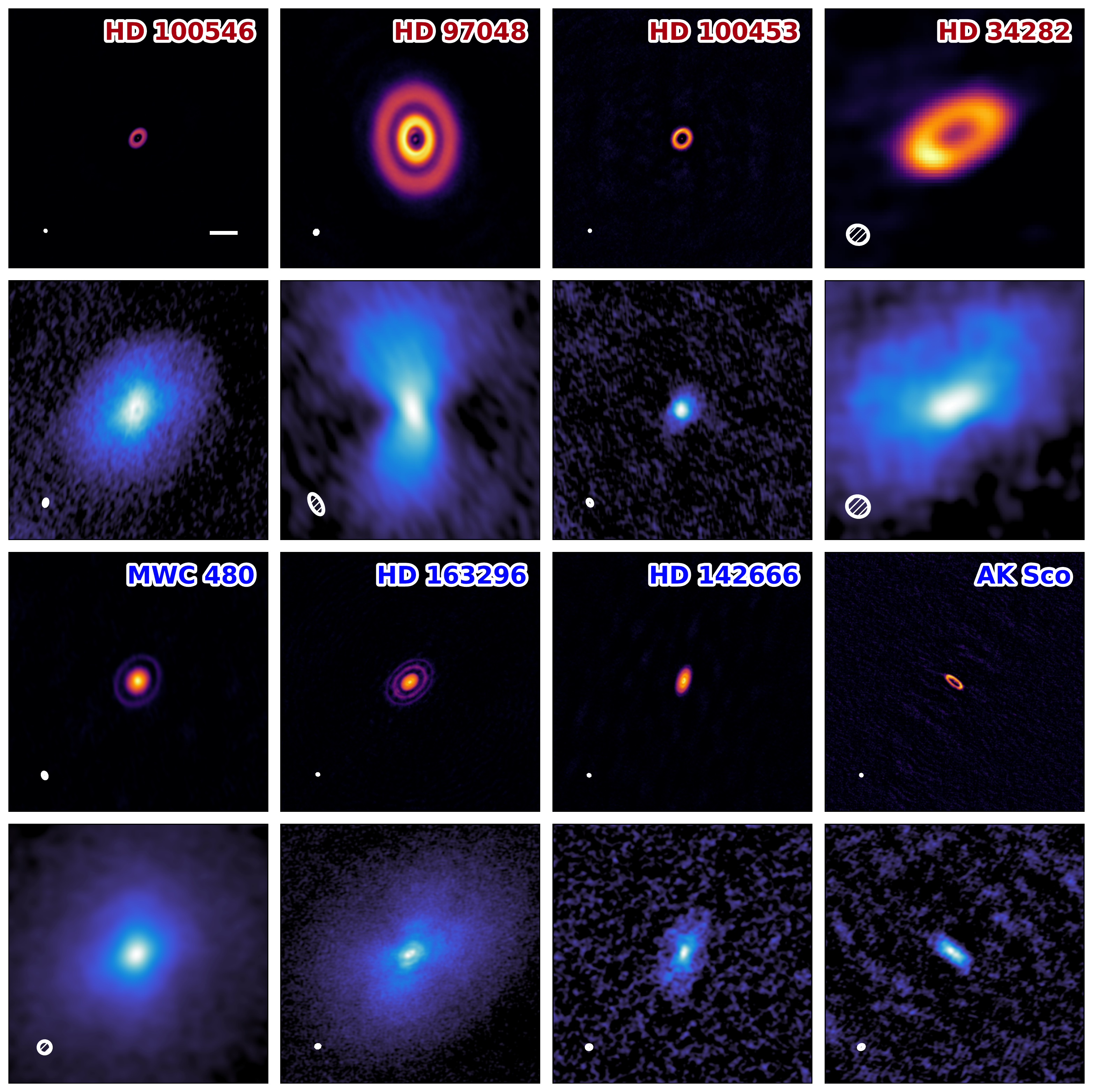}
    \caption{Continuum and $^{12}$CO velocity integrated (moment 0) maps of the group I disks (top two rows) and group II disks (bottom two rows). Each image is 1000$\times$1000~au in size, additionally a bar of 100~au in size is shown in the top left panel. In the bottom left corner of each image the size of the beam is shown. The moment 0 maps use a 3$\sigma$ clip and only include channels with disk emission. To make the outer regions of the disks better visible, a power-law normalization is used}
    \label{fig:gallery}
\end{figure*}

Herbig Ae/Be stars \citep[e.g.,][]{Herbig1960, Waters1998} are pre-main sequence stars of intermediate mass and spectral type between B and mid-F. The disks around Herbig stars (hereafter Herbig disks) have been found to be more massive than those around T-Tauri stars \citep{Acke2004, vanderMarel2021, Stapper2022}. This could explain why giant planets are more common around intermediate mass main-sequence stars \citep[e.g.,][]{Johnson2007, Fulton2021}. The potential of forming a giant planet may be linked to the group I and II spectral energy distribution (SED) characterization of Herbig disks \citep{Stapper2022}. These two groups are based on the shape of the SED, specifically on the infrared (IR) excess \citep{Meeus2001, Acke2009}. While group II SEDs can be fit with a single (power) law across the near--far infrared wavelength range, group I SEDs require an additional blackbody component that dominates the far-infrared emission. These differences were interpreted as the group I disks being flaring disks (or vertically extended, i.e., increasing height with radius), while the group II disks are flat (i.e., constant height with radius) or self-shadowed \citep{Meeus2001}.

Originally, an evolutionary sequence from group I to group II was hypothesized, with grain growth and settling reducing the far-infrared emission \citep{Dullemond2004a, Dullemond2004b, Dullemond2005}. However, their evolution has been shown to be more complicated. It was found that most, if not all, Herbig group I disks have inner cavities \citep{Honda2012, Maaskant2013}, which gave rise to the idea that the flux increase at longer wavelengths is due to an irradiated inner cavity wall. Consequently, an evolution from group I to group II was discarded and instead, it was proposed that group II objects might eventually evolve into group I objects due to the creation of an inner cavity \citep{Maaskant2013, Menu2015}.

However, the lack of scattered light from group II objects suggest that these still may be flat, self-shadowed disks \citep[e.g.][]{Garufi2017, Garufi2022}. This lead \citet{Garufi2017} to propose that group II disks can either be self-shadowed large disks, or small compact disks. This is in contrast to the group I disks, all of which are large and bright in both scattered light \citep{Garufi2014} and mm-observations \citep{Stapper2022}. This leads to the question of whether similar differences in vertical disk height are also present in the gas.

In recent years, the high velocity and spatial resolution of the Atacama Large Millimeter/submillimeter Array (ALMA) has allowed for the characterization of the vertical structure of mid-inclination protoplanetary disks \citep[e.g.,][]{Pinte2018, Rich2021, Law2021, Law2022, PanequeCarreno2021, PanequeCarreno2022, PanequeCarreno2022b} enlarging the number of disks accessible in addition to edge-on disks \citep[e.g.,][]{Podio2020, Villenave2020}. The emission heights of different molecules \citep[e.g.,][]{Law2021}, and asymmetries in the vertical emission of CO in a disk \citep{PanequeCarreno2021} have been found. In this paper, we apply this technique to determine the emission heights of $^{12}$CO for both group I and group II disks and see if there are any differences present between the two groups.

Section \ref{sec:sample_selection_and_data_reduction} presents the targets, how they were selected and imaged. Section \ref{subsec:disk_heights} shows the extracted disk heights and Section \ref{subsec:disk_temperatures} the temperature maps of each disk. The results are discussed in the context of group I vs group II in \S\ref{subsec:groupI_vs_groupII} and the age related to the very flat disks in \S\ref{subsec:old_disks}. Lastly, in Section \ref{sec:conclusion} the conclusions are summarized.

\begin{table*}[t]
\small
\caption{Data and stellar parameters of each Herbig disk.}
\begin{tabular}{c l c c c c c c c c c c}
\hline\hline
\makecell{Group \\ \hspace{1mm}} & \makecell{Herbig disk \\ \hspace{1mm}} & \makecell{Vel. res. \\ (km~s$^{-1}$)} & \makecell{Sp. res. \\ ($''$)} & \makecell{rms \\ (mJy beam$^{-1}$)} & \makecell{Project ID \\  \hspace{1mm}} & \makecell{Dist. \\ (pc)} & \makecell{M$_\star$ \\ ($M_\odot$)} & \makecell{$L_\star$ \\ (L$_\odot$)} & \makecell{Age \\ (Myr)} & \makecell{Inc. \\ ($\degree$)} & \makecell{PA \\ ($\degree$)} \\ \hline
I     & HD~34282    & 0.2  & 0.27 & 4.6 & 2015.1.00192.S & 306.5 & <1.9 & 14.5 & <20 & 60 & 117  \\
      & HD~97048    & 0.3  & 0.46 & 4.2 & 2015.1.00192.S & 184.1 &  2.8 & 64.6 &   4 & 41 &   3  \\
      & HD~100453   & 0.3  & 0.25 & 3.5 & 2015.1.00192.S & 103.6 &  1.6 &  6.2 &  19 & 30 & 140  \\
      & HD~100546   & 0.2  & 0.24 & 4.2 & 2016.1.00344.S & 108.0 &  2.1 & 21.9 &   8 & 43 & 139  \\ \hline
II    & AK~Sco      & 0.3  & 0.15 & 2.5 & 2016.1.00204.S & 139.2 &  1.7 &  5.6 &   8 & 109 &  51  \\
      & HD~142666   & 0.35 & 0.13 & 1.6 & 2016.1.00484.L & 145.5 &  1.8 & 13.5 &   9 &  62 & 162  \\
      & HD~163296   & 0.2  & 0.14 & 0.6 & 2018.1.01055.L & 100.6 &  1.9 & 15.5 &  10 &  46 & 312  \\
      & MWC~480     & 0.2  & 0.3  & 1.2 & 2018.1.01055.L & 155.2 &  1.9 & 16.6 &   8 & -32 & 328  \\ \hline
\end{tabular}\\
\textbf{Notes:} The stellar parameters are taken from \citet{GuzmanDiaz2021}. The inclinations and position angles are taken from: AK~Sco: \citet{Czekala2015}, HD~142666: \citet{Huang2018}, HD~163296: \citet{Izquierdo2022}, MWC~480: \citet{Teague2021}, HD~34282: \citet{vanderPlas2017b}, HD~100453: \citet{Rosotti2020}, HD~97048: \citet{Walsh2016}, HD~100546: \citet{Pineda2019}.
\label{tab:params}
\end{table*}

\begin{figure*}[h]
    \centering
    \includegraphics[width=0.8\textwidth]{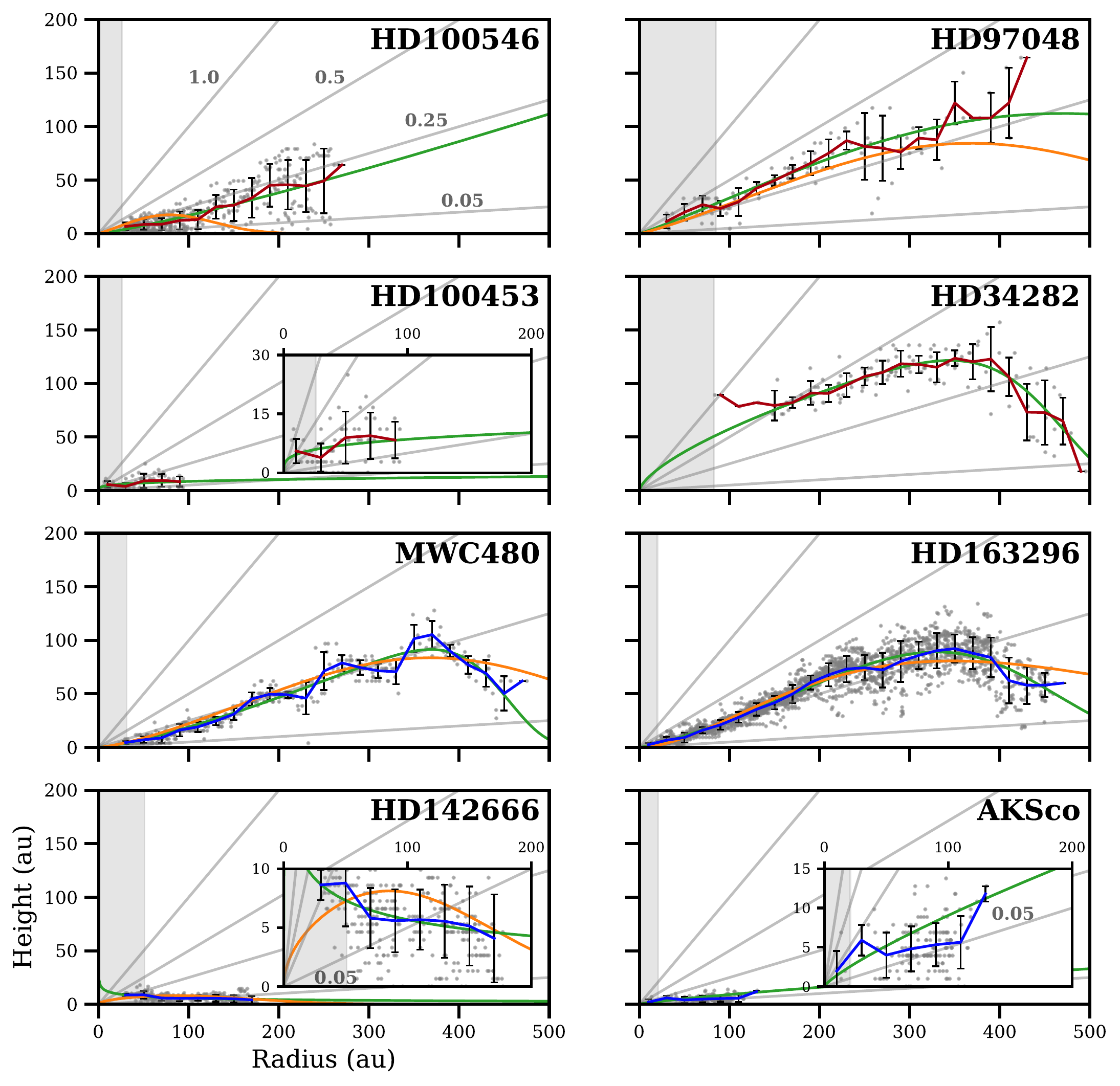}
    \caption{Height profiles of the group I disks (top four panels) and group II disks (bottom four panels). The gray lines indicate the 1.0, 0.5, 0.25 and 0.05 height to radius ratios. The gray scatter is the extracted points from the channel maps and the line shows the mean value of the scatter in bins of 20~au in size. The errorbars indicate the standard deviation of the scatter in each bin. The fitted profiles are shown as the green lines and are compared to the orange profiles from \citet{Law2021} for HD~163296 and MWC~480, \citet{Law2022} for HD~142666 and HD~100546, and \citet{Rich2021} for HD~97048. The gray shaded regions indicate the size of the major axis of the beam.}
    \label{fig:height_profiles}
\end{figure*}

\section{Sample selection and Data reduction}
\label{sec:sample_selection_and_data_reduction}
In recent years many Herbig disks have been observed with ALMA \citep[for a recent compilation, see][]{Stapper2022}. Some of these data are deep, high resolution observations. These data were either included in ALMA large programs such as MAPS \citep{Oberg2021} and DSHARP \citep{Andrews2018b}, or have been studied separately. Our sample was selected based on the available data in the ALMA archive\footnote{\url{https://almascience.eso.org/asax/}} \citep{Stapper2022}. To be able to determine the emission heights of a disk, the data and the disk need to meet several requirements. First, an inclination of at least $30\degree$ is needed to extract the surface emitting heights reliably \citep{Law2021}. Second, high enough velocity resolution is necessary to adequately trace the isovelocity curves in the selected channels. In general, a velocity resolution of at least $\sim$0.3~km~s$^{-1}$ is necessary. Lastly, sufficient spatial resolution is necessary to distinguish both the near-side and far-side of the disk \citep[for systematics see,][]{Pinte2018, PanequeCarreno2022}. For our sources, this comes down to $\sim30$~au. To properly trace the vertical extent of the Herbig disks, we will use $^{12}$CO $J=2-1$ observations. This leaves us with eight total data sets, four group I and four group II disks, for which these requirements are met, see Table \ref{tab:params}. The SEDs of these disks are shown in Appendix \ref{app:SEDs}, which clearly show the distinguishing features of both groups. We note that this technique of determining the vertical extent of the disk traces the $\tau=1$ line rather than the scale height of the gas. \citet{PanequeCarreno2022b} has shown that the ratio between the traced emission height and the gas scale height is generally a factor of 2-5.

To obtain well-defined upper surfaces, $^{12}$CO observations towards the eight disks are used. For HD~163296 and MWC~480, the imaged $^{12}$CO data sets from the MAPS large program were used (see \citealt{Czekala2021} for the specifics on the imaging). All other data were imaged using the \texttt{Common Astronomy Software Applications} (CASA) application version 5.8.0 \citep{McMullin2007}. For all data sets either the data were binned by a factor of two in velocity to increase the sensitivity or the native velocity resolution was used. After subtracting the continuum using \texttt{uvcontsub} (which was not done for the temperature maps shown in Section \ref{subsec:disk_temperatures}), the data were imaged using the \texttt{multiscale} algorithm. The used scales were 0 (point source), 1, 2, 5, 10 and 15 times the size of the beam in pixels ($\sim$5 pixels). A Briggs robust weighting of 0.5 was used. For HD~142666 a $uv$-taper of 0.08$''$ was applied to increase the beamsize for better extraction of the emission surface heights. These steps resulted in the image parameters listed in Table \ref{tab:params}. The velocity integrated maps of $^{12}$CO can be found in Fig. \ref{fig:gallery} together with the Band 6 or 7 continuum images of each disk.

To obtain the emitting surface heights of the $^{12}$CO emission, the same technique as set out in \citet{Pinte2018} is used. Using geometric relations and assuming Keplerian rotation, the emission height can be retrieved. We use the same implementation of this technique as \citet{PanequeCarreno2021, PanequeCarreno2022, PanequeCarreno2022b}, in which the distinction between the upper and lower emission surfaces is done visually with the use of hand-drawn masks using \texttt{ALFAHOR} \citep[ALgorithm For Accurate H/R,][]{PanequeCarreno2022b}; see Appendix \ref{app:channel_maps}. This limits contamination between the different surfaces, resulting in a cleaner retrieval of the emission surfaces. The extracted data points are averaged in bins of 20~au in size, with an uncertainty corresponding to the standard deviation of the data points in that bin. The distances used and other stellar parameters can be found in Table \ref{tab:params}. The resulting hand-drawn masks and extracted points are shown in the figures of Appendix \ref{app:channel_maps}.

An exponentially tapered power law is fitted to each profile, to be able to easily compare between disks and other works. We follow \citet{Law2021} in using the following expression:

\begin{equation}
    z(r) = z_0 \times \left(\frac{r}{1''}\right)^\phi \times \exp\left(-\left[ \frac{r}{r_\text{taper}} \right]^\psi\right),
    \label{eq:power_law}
\end{equation}

\noindent
where $z$ is the vertical height of the surface emission, $r$ is the radius of the profile, and $z0$ and $r_\text{taper}$ are related to the size of the disk in the vertical and radial direction respectively. All units are in arcseconds. For fitting Eq. (\ref{eq:power_law}), we use the \texttt{curve\_fit} function from \texttt{SciPy} \citep{2020SciPy-NMeth} to do a non-linear least-squares fit to the retrieved binned emission surfaces. Bins with only one point are excluded.
Lastly, for additional visual aide in identifying if disks are vertically extended or flat, Appendix \ref{app:velocity_maps} presents the velocity maps of each disk.

\begin{figure*}
    \centering
    \includegraphics[width=\textwidth]{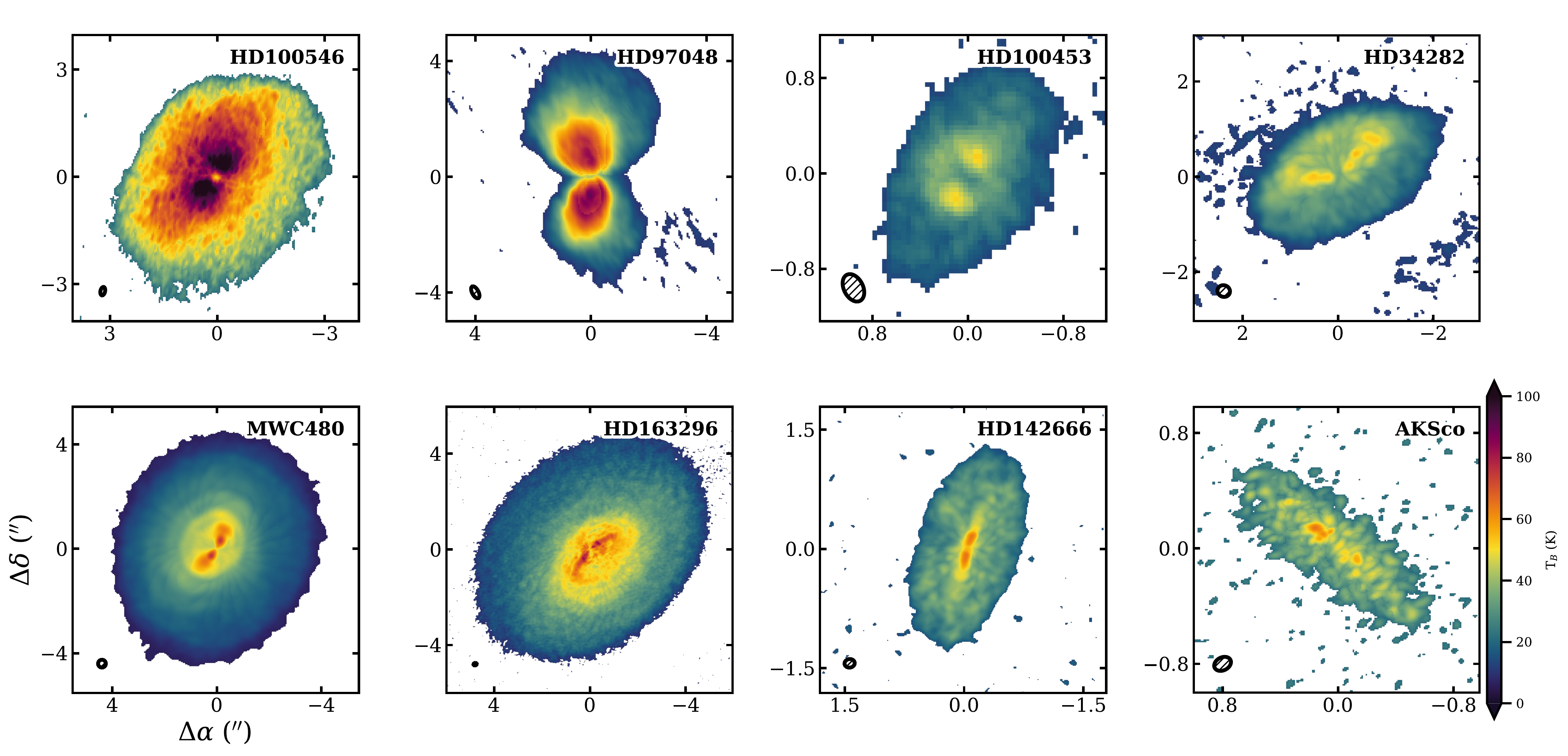}
    \caption{Temperature maps of the eight disks, determined using the Planck law. The top and bottom rows show the group I and II disks respectively. Each image is clipped at an SNR of 3. The size of the beam is shown in the bottom left corner.}
    \label{fig:temperature}
\end{figure*}

\section{Results}
\label{sec:results}
\subsection{Disk heights}
\label{subsec:disk_heights}
Both the continuum and $^{12}$CO line observations presented in Figure \ref{fig:gallery} show already a large variety in structures in and sizes of the disks in this work. Both HD~97048 and HD~34282 show both a large dust and gas disk extent. The dark regions in the $^{12}$CO observations of HD~97048 is due to foreground cloud absorption. For the group I disks, all disks except HD~100453 are large in gas. For the group II disks both HD~142666 and AK~Sco are smaller compared to the other disks. The continuum of all disks show substructure, half of the disks show a single dust ring while the others show multiple rings.

Figure \ref{fig:height_profiles} presents the extracted height profiles for the eight disks, in which the top four panels show the group I disks and the bottom four panels show the group II disks. In general the heights of most of the disks are at a $z/r\sim0.25$. All group I disks have a $z/r$ of at least 0.25, either in large parts of the disk such as HD~100546 or HD~97048 or at small radii only (HD~100453). MWC~480 and HD~163296, both belong to group II and are vertically very similar to the group I disks. Both disks have a $z/r\sim0.25$ and have gas disk sizes of around 500~au. On the contrary, AK~Sco and HD~142666 are both flat and relatively small, only going out to 200~au. We will now discuss each disk separately.

HD~34282, while having similar or worse spatial resolution compared to the other disks, is limited by the resolution of the data due to being the farthest away source \citep[306.5~pc, see][or Table \ref{tab:params}]{GuzmanDiaz2021}. For the inner 100~au, the far and near side of the disk cannot be separated from each other, making it impossible to sample the emission height at these radii (see Fig. \ref{fig:HD34282_channel_maps}). The part that is well sampled shows a vertically extended disk, going as high as $z/r\sim0.5$. In the case of HD~34282, the tapered power-law fit shows that at small radii the disk could go above $z/r\sim0.5$, which is higher than what is found in most disks \citep[e.g.,][]{Law2022}.

\begin{figure*}[t]
    \centering
    \includegraphics[width=\textwidth]{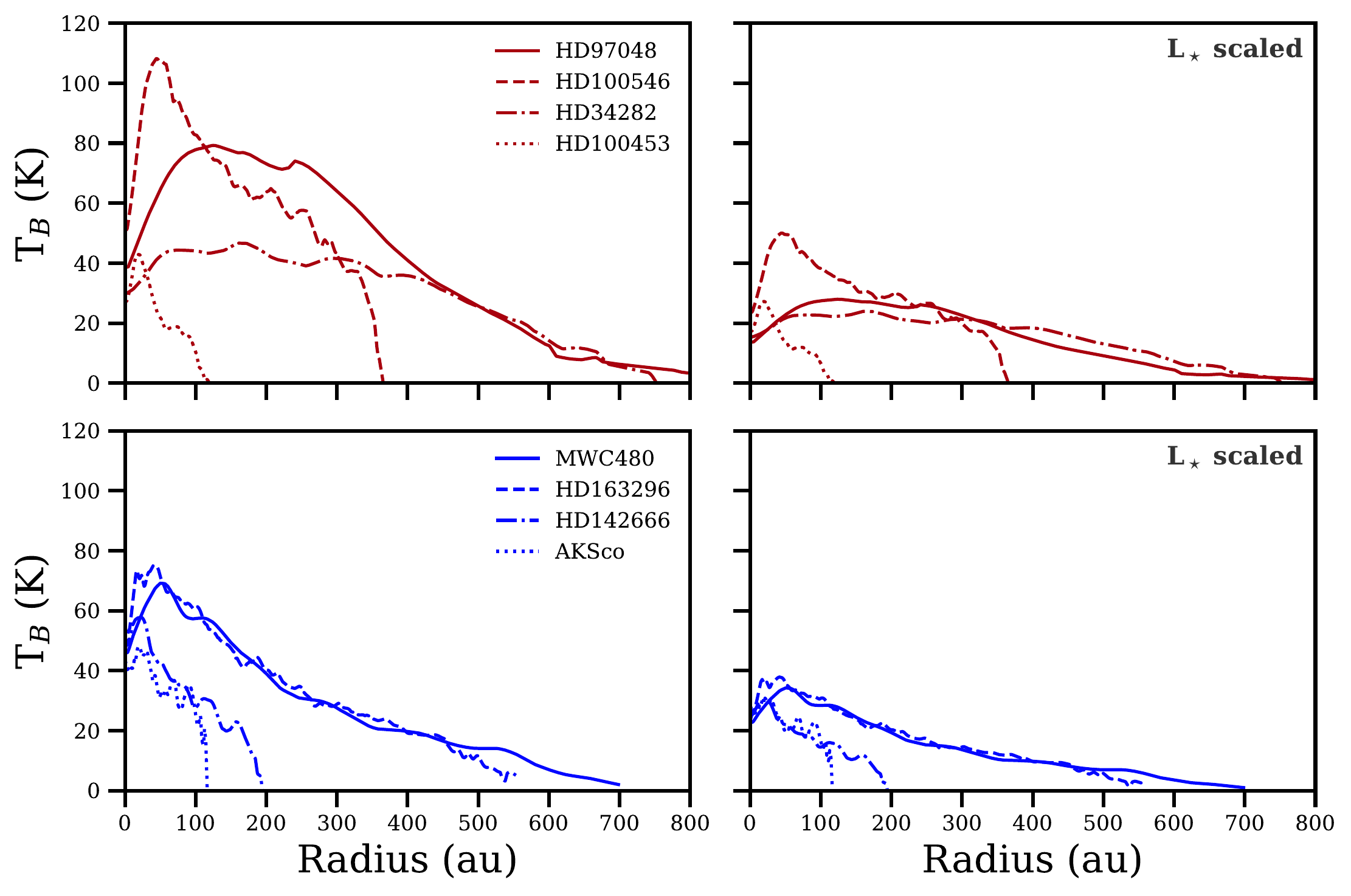}
    \caption{Radial cuts of the temperature maps in Fig. \ref{fig:temperature} along the major axes of the disks. Colored by group I (red, top row) and group II (blue, bottom row). The legend is ordered by stellar luminosity. The right panels show the temperature scaled by the stellar luminosity as $T\sim(L_\star/L_\odot)^{1/4}$ from \citet{Andrews2013}.}
    \label{fig:radial_cut}
\end{figure*}

HD~97048 is also a high vertically extended disk in our sample with a $z/r$ out to 0.33. Although we do not probe the turnover at large radii as seen in many disks, we find that the emission originates from higher $z/r$ regions in comparison to what \citet{Rich2021} found. While the inner region is again limited by the spatial resolution of the data, the profile is above a $z/r$ of 0.25 at these regions close to the midplane. The $z/r$ values close to the midplane of the best fit agree well with those of \citet{Rich2021}, while the outer regions mainly contribute to the difference between the two fits. The data lack samples at lower velocities close to the system velocity (see Fig. \ref{fig:mom1} in Appendix \ref{app:velocity_maps}) due to foreground cloud absorption, which reduces the number of sampled heights in Fig. \ref{fig:height_profiles}. For HD~100546 we are able to probe the emission heights to larger radii than what was found by \citet{Law2022}. However, no clear turnover is detected at these larger radii up to where we sample the surface. The disk clearly follows a $z/r$ $\sim0.25$.

Fitting the exponentially-tapered power law to the height profiles of MWC~480 and HD~163296, we find a steeper drop off at large radii compared to what \citet{Law2021} find. The main difference here is due to being able to better separate the lower and upper emission surfaces by drawing masks, which would otherwise contaminate the overall inferred profile. At closer separations, our relations agree with what \citet{Law2021} found, while due to a better sampling at larger separations the relations start to differ at larger radii, finding a steeper drop.

For HD~142666 we find a very flat disk, similar to the disk height found by \citet{Law2022}, but there are differences at small radii. In addition to HD~142666, we find a second very flat disk: AK~Sco. Both of these disks are also the smallest disks in the sample of this work.

\subsection{Disk temperatures}
\label{subsec:disk_temperatures}
Is the height of the emitting layer, and the larger amount of intercepted stellar light, reflected in the temperature of the emitting gas?

In Fig. \ref{fig:temperature} the temperature maps of the disks are presented. These maps were made with \texttt{bettermoments} using a clip of 3$\times$SNR and the full Planck expression. A radial cut along the disks major axes following the projected fitted height profiles of these temperature maps can be found in the left column of Fig. \ref{fig:radial_cut}. Each disk shows a decrease in temperature close to the star due to beam dilution, as the emitting region shrinks and no longer fills the beam \citep[e.g.,][]{Leemker2022}. Also, for HD~142666, the relatively low velocity resolution can lower the inferred temperature by underresolving the line. Both Fig. \ref{fig:temperature} and \ref{fig:radial_cut} show large differences between temperatures of individual disks. However, no clear trend is present between the group I and group II sources. Disks from either group can be found with very similar temperature profiles. For example HD~100453/AK~Sco/HD~142666 and HD~34282/HD~163296/MWC~480. The only clear outliers are HD~100546 and HD~97048, which are much warmer in the region out to 450~au compared to the other disks.

Are these differences due to their relative vertical extent? The right column of Fig. \ref{fig:radial_cut} shows radial cuts of the temperature maps where the temperature is scaled by the stellar luminosity as $\sim(L_\star/L_\odot)^{1/4}$ \citep[following the dust temperature scaling presented in][]{Andrews2013}. This scaling removes the effect of the central star on the temperature of the disk. As presented in Table \ref{tab:params}, both HD~100546 and HD~97048 are the most luminous stars in our sample, which have increased the temperatures significantly relative to the less luminous stars. When this scaling is applied their relative temperatures become noticeably more similar. Especially the temperature profile of HD~97048 becomes almost identical to the profiles of HD~34282, MWC~480 and HD~163296. Consequently, their temperatures are as one would expect for their luminosity and size. Hence, we do not find a difference in temperature between the group I and group II sources in spite of differences in vertical extent.

\citet{Fedele2016} modeled Herbig~Ae disks based on Herschel/HIFI high-J CO line profiles and determined the radial gas temperature structure for, among others, different vertically extended disks. They find that for $z/r$ between 0.01 and 0.3 the gas temperature is independent of the vertical extent of the disk: the differences only start in higher layers probed by (very) high-J CO lines. Given that our $^{12}$CO emission surfaces are from $z/r\sim0.3$ and lower, our results fall in line with these models.

\section{Discussion}
\label{sec:discussion}

\subsection{Group I vs Group II}
\label{subsec:groupI_vs_groupII}

Originally, the different Herbig disk groups were interpreted as group I being a flaring disk and group II being a flat or self-shadowed disk with an evolution from group I to group II via dust settling. As outlined in Section \ref{sec:introduction}, since then this view has become more complicated. Based on scattered light imaging, \citet{Garufi2017} proposed that group II disks can be divided in objects with large, shadowed disks and objects with small disks. Both result in a low far-infrared flux. The shadowed disks could then evolve into group I disks by creating an inner cavity, and thus removing the part of the disk casting the shadow. Recent studies have shown a high occurrence rate of cavities in group I disks in both gas and dust \citep[e.g.,][]{Menu2015, vanderPlas2015}.

For most group I disks we find large vertically extended disks. For the disks in group II, there is a clear difference between two sets of disks: MWC~480 and HD~163296, versus HD~142666 and AK~Sco. The former two disks are indistinguishable from the group I disks in our sample, both regarding their size and the vertical extent of the disk while the latter are more compact and flat.

These two sets of disks coincide with the proposed distinction between compact and self-shadowed group II disks by \citet{Garufi2017}. The analogous height profiles of MWC~480 and HD~163296 compared to the group I disks suggest that these are precursors of the group I disks, as \citet{Garufi2017} suggest for HD~163296. These disks are shadowed due to an inner disk, keeping the outer regions cool, which results in a group II SED. Once a cavity forms, the shadowing disappears and a group I disk is formed. Throughout this evolution, the height profile of the disk stays the same. The puzzling aspect is that a higher temperature should result in a more vertically extended disk, which is not observed. So there may be a (unrelated) trend occurring at the same time that results in a flatter disk, counteracting the effect of the larger illumination. The two flatter disks are also the more compact disks in our sample. This also creates a group II SED.

Hence, two types of group II disks can be distinguished: On the one hand those disks that will eventually evolve in group I disks, with very similar radial and vertical structure but where the inner disk shadows the outer disk. And on the other hand disks that are very flat, and never developed an inner cavity that exposes the outer disk to heating.

We note that no differences between the vertically extended but shadowed and vertically flat disks are seen in the SEDs of the group II disks (see Appendix \ref{app:SEDs}). As shown in \citet{Garufi2017, Garufi2022}, in general multiple tracers are necessary to fully characterize a Herbig disk, one of which is the ability to spatially resolve the disk.

\subsection{Old flat group II disks}
\label{subsec:old_disks}
Both AK~Sco and HD~142666 are very flat disks compared to the other disks in our sample. What could cause such a flat disk?

The ages of these disks are not well determined and have a significant range in possible values. For instance, in Table \ref{tab:params} an age of 7.8~Myr is cited for AK~Sco \citep{GuzmanDiaz2021}. However, others give ranges of values from lower limits of 12~Myr \citep{Garufi2022} to 18~Myr \citep{Czekala2015}.  While not present in this work due to too low spectral resolution observations, HD~9672 (or 49~Ceti) is a group II Herbig disk but is also considered to be a debris disk \citep[e.g.,][]{Moor2019}. The secondary dust in the debris disk, which is expected to have a low scale height, could cause the group II classification. At the advanced age of these disks the PAHs and small dust grains in the higher regions of the disk may have been removed. This in turn lowers the gas temperature and decreases the gas scale height. Hence for AK~Sco and HD~142666, while not being debris disks, significantly more evolution (dust settling and thus a decrease of the gas scale height) might have happened in these flat disks than one would assume based on the ages mentioned in Table \ref{tab:params}.

\section{Conclusion}
\label{sec:conclusion}
In this work we determined the emission heights of eight Herbig disks, four group I (HD~100546, HD~97048, HD~100453 and HD~34282) and four group II disks (MWC~480, HD~163296, HD~142666 and AK~Sco) following the classification of \citet{Meeus2001}. With these emission heights, we test the interpretation that group I disks are \textbf{vertically extended} irradiated disks, and group II disks are self-shadowed or flat disks. The following conclusions are made:

\begin{enumerate}
    \item All but one of the group I disks are large (>200~au) and have $z/r\sim0.25$. The exception is HD~100453, which has the same $z/r$ as the other group I disks, but is traced to smaller radii.
    \item Two of the group II disks (MWC~480 and HD~163296) have indistinguishable emission height profiles compared to those of the group I disks.
    \item We find two very flat disks among the group II disks (HD~142666 and AK~Sco). Their emission heights are below 10~au over the full extent of the disk traced (out to 200~au).
    \item The temperatures reveal no significant differences between the disks when scaled based on the luminosity of the star, in spite of differences in vertical extent.
    \item Our findings agree with the proposed scenario of \citet{Garufi2017} where some group II disks are self-shadowed and will evolve into a group I disk by forming a cavity which causes the outer disk to be irradiated (MWC~480 and HD~163296), and other group II disks being small and flat (AK~Sco and HD~142666).
    \item No significant differences are present between the SEDs of the flat and the vertically extended group II disks. Hence, resolved observations play a key role in fully characterizing the different Herbig disk populations.
\end{enumerate}

\noindent The small source sample of only four objects in each SED group with available ALMA data of sufficient quality shows that more observations of sufficient $S/N$ and resolution, both spatial and kinematic, are needed. Future studies with a larger sample must be done to place the dichotomy in gas disk heights reported by us on a firm statistical footing.

\begin{acknowledgements}
The research of LMS is supported by the Netherlands Research School for Astronomy (NOVA). This paper makes use of the following ALMA data: 2015.1.00192.S, 2016.1.00344.S, 2016.1.00204.S, 2016.1.00484.L, 2018.1.01055.L. ALMA is a partnership of ESO (representing its member states), NSF (USA) and NINS (Japan), together with NRC (Canada), MOST and ASIAA (Taiwan), and KASI (Republic of Korea), in cooperation with the Republic of Chile. The Joint ALMA Observatory is operated by ESO, AUI/NRAO and NAOJ. This work makes use of the following software: The Common Astronomy Software Applications (CASA) package \citep{McMullin2007}, Python version 3.9, alfahor \citep{PanequeCarreno2022b}, astropy \citep{astropy2013, astropy2018}, bettermoments \citep{Teague2018}, matplotlib \citep{Hunter2007}, numpy \citep{Harris2020} and scipy \citep{2020SciPy-NMeth}. Our thanks goes to the European ARC node in the Netherlands (ALLEGRO) for their support with the calibration and imaging of the data. Lastly,  we thank the referee for their insightful comments which have improved this paper.

\end{acknowledgements}

\bibliographystyle{aa}
\bibliography{references.bib}

\begin{appendix}
\section{Spectral Energy Distributions}
\label{app:SEDs}
\begin{figure*}[b]
    \centering
    \includegraphics[width=\textwidth]{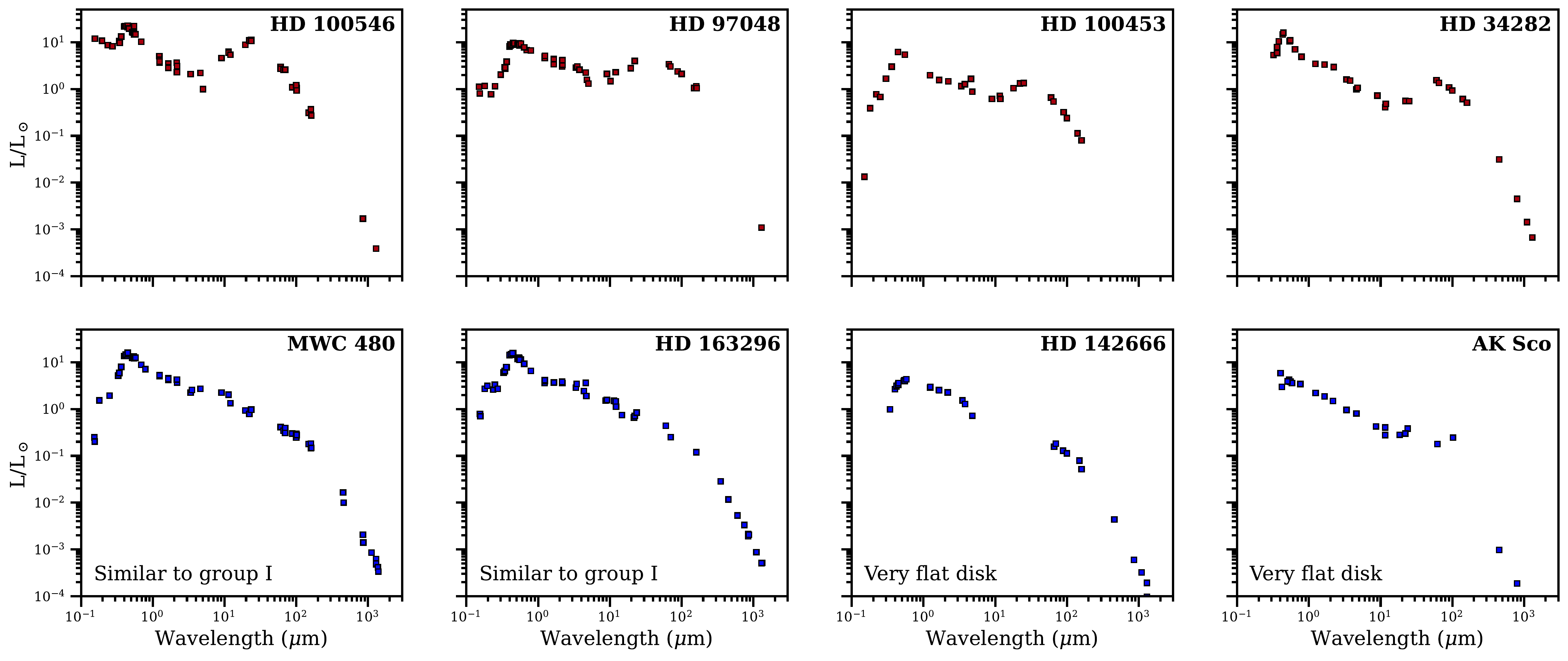}
    \caption{Spectral energy distributions (SEDs) of the eight disks. The top and bottom rows show the SEDs of the group I and group II sources respectively. The SEDs of HD~100546, HD~97048, MWC~480, HD~163296 and HD~142666 are from the DIANA project \citep{Woitke2019}. The SEDs of HD~100453 and HD~34282 are from \citet{Khalafinejad2016}. Lastly, the SED of AK~Sco is compiled from the following sources: \citet{Hipparcos1997, Hindsley1994, Zacharias2004, GAIADR3, Ishihara2010, Cutri2012, Jensen1996}. The SEDs of HD~100453, HD~34282 and AK~Sco have been deredened using the \texttt{Astropy} affiliated package \texttt{dust\_extinction} with $R_V=3.1$ \citep{Bessell1979}, and a visual extinction ($A_V$) from \citet{GuzmanDiaz2021}.}
    \label{fig:SEDs}
\end{figure*}

Figure \ref{fig:SEDs} presents the SEDs of the eight Herbig disks. The main distinction between the group I and group II disks is clear: group I disks have extra emission at mid-far infrared wavelengths, while group II disks only decrease in emission with increasing wavelength.

\section{Channel maps}
\label{app:channel_maps}
\begin{figure*}
    \centering
    \includegraphics[width=\textwidth]{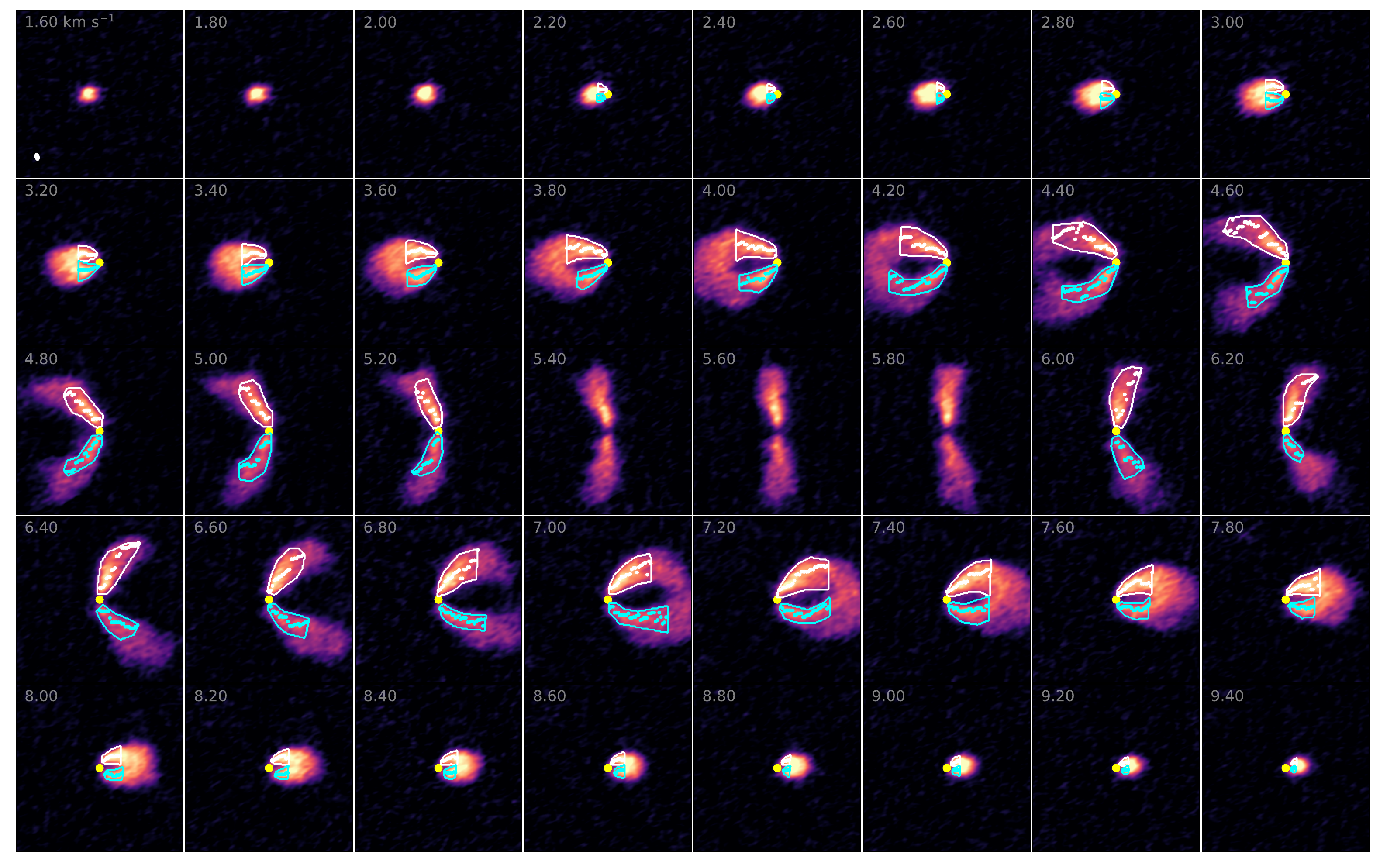}
    \caption{$^{12}$CO channel maps of HD~100546.}
    \label{fig:HD100546_channel_maps}
\end{figure*}

\begin{figure*}
    \centering
    \includegraphics[width=\textwidth]{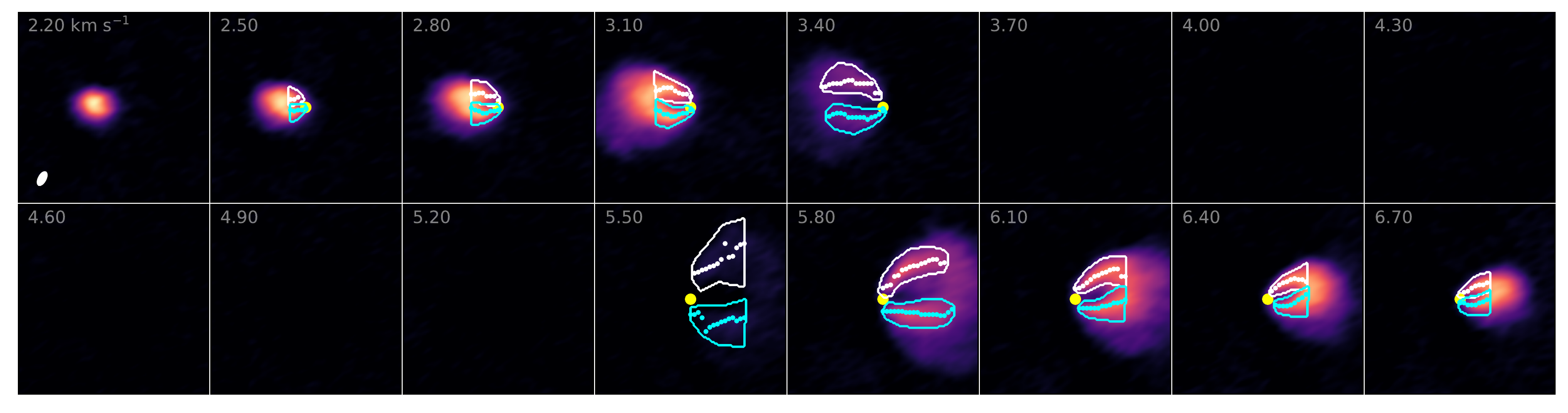}
    \caption{$^{12}$CO channel maps of HD~97048.}
    \label{fig:HD97048_channel_maps}
\end{figure*}

\begin{figure*}
    \centering
    \includegraphics[width=\textwidth]{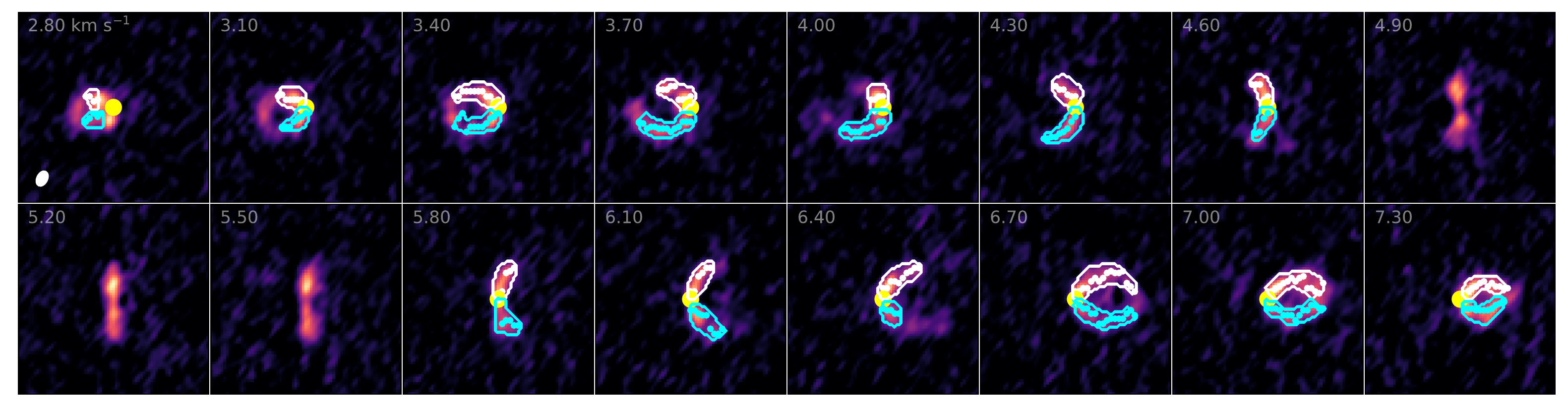}
    \caption{$^{12}$CO channel maps of HD~100453.}
    \label{fig:HD100453_channel_maps}
\end{figure*}

\begin{figure*}
    \centering
    \includegraphics[width=\textwidth]{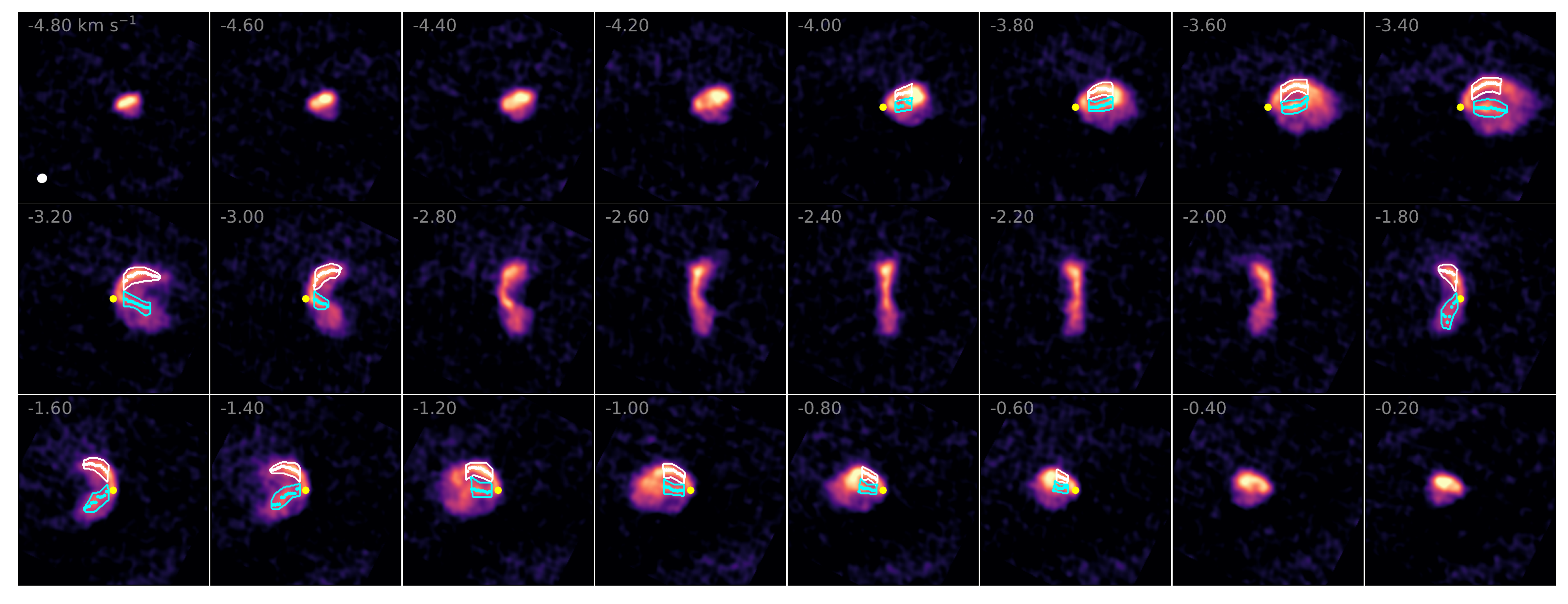}
    \caption{$^{12}$CO channel maps of HD~34282.}
    \label{fig:HD34282_channel_maps}
\end{figure*}

\begin{figure*}
    \centering
    \includegraphics[width=\textwidth]{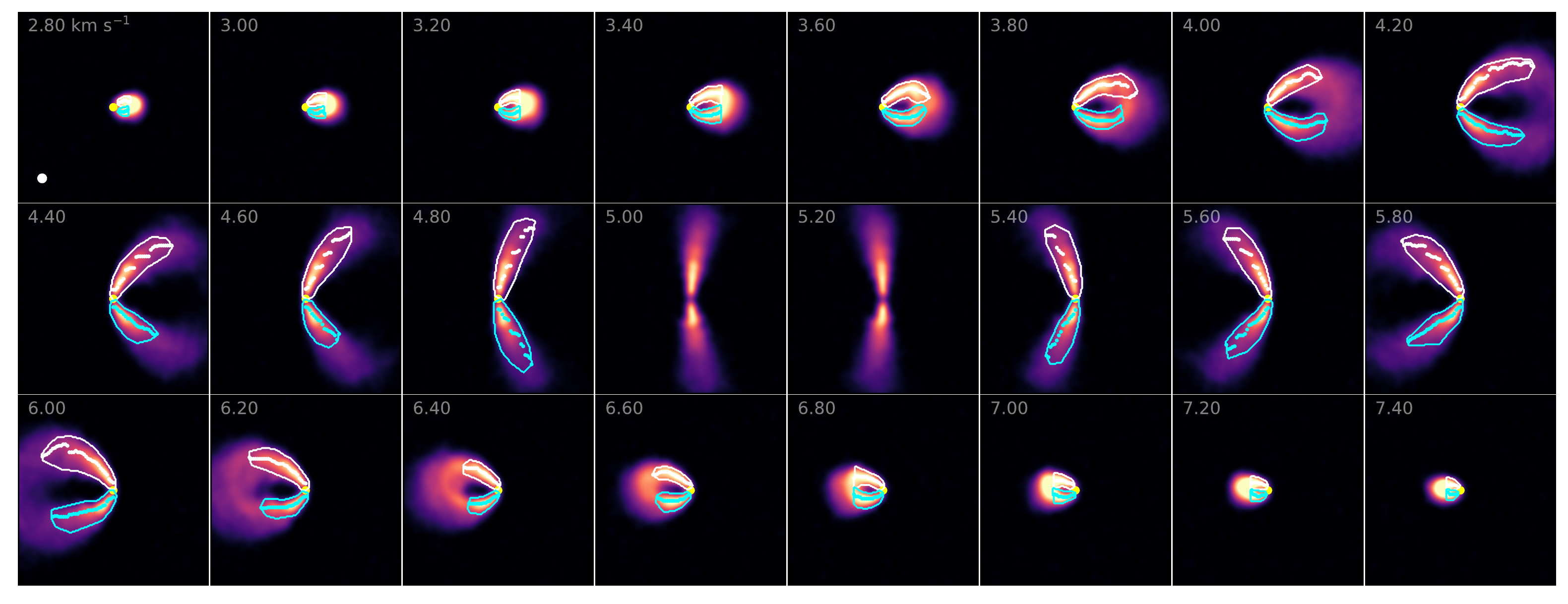}
    \caption{$^{12}$CO channel maps of MWC~480.}
    \label{fig:MWC480_channel_maps}
\end{figure*}

\begin{figure*}
    \centering
    \includegraphics[width=\textwidth]{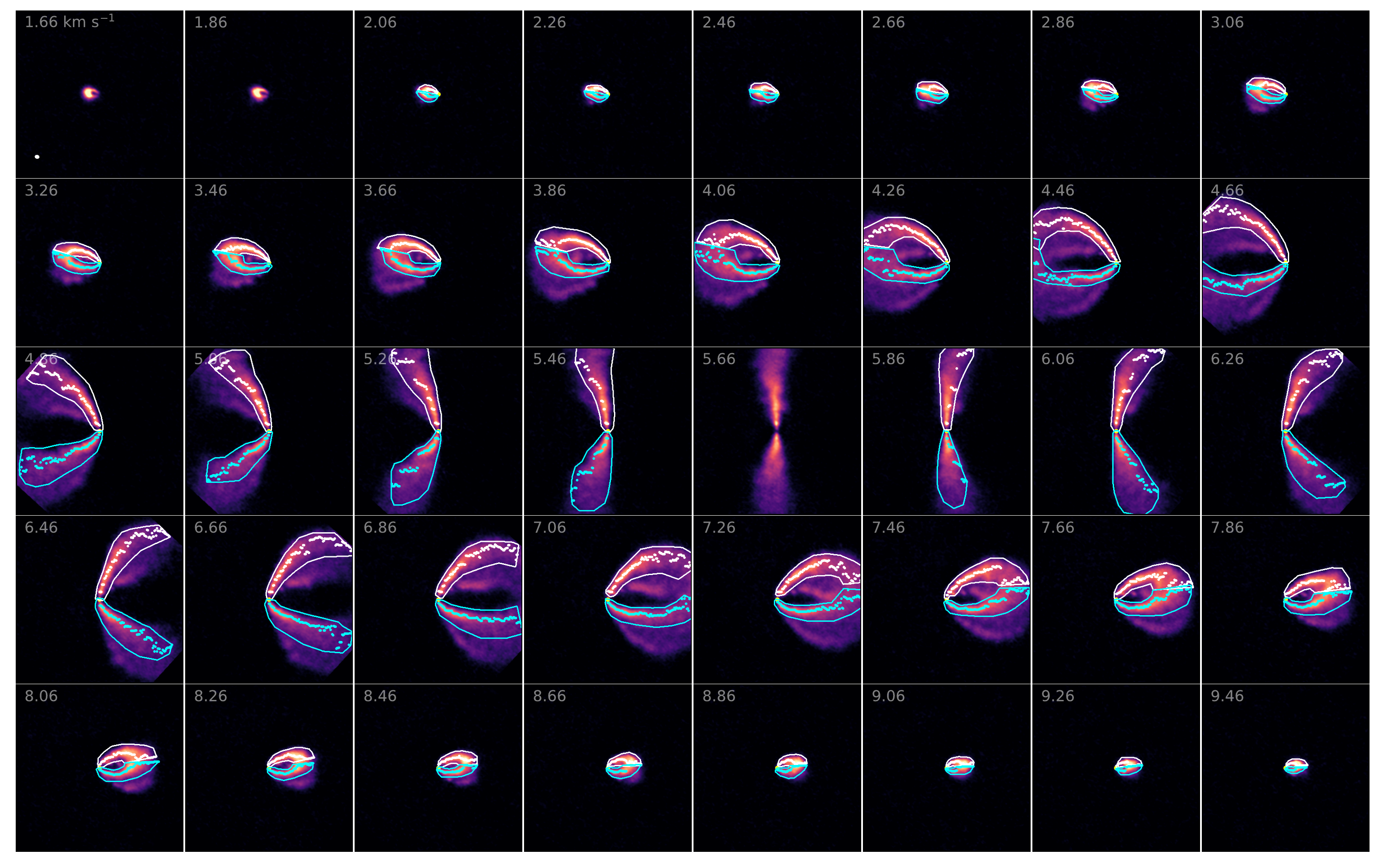}
    \caption{$^{12}$CO channel maps of HD~163296. }
    \label{fig:HD163296_channel_maps}
\end{figure*}

\begin{figure*}
    \centering
    \includegraphics[width=\textwidth]{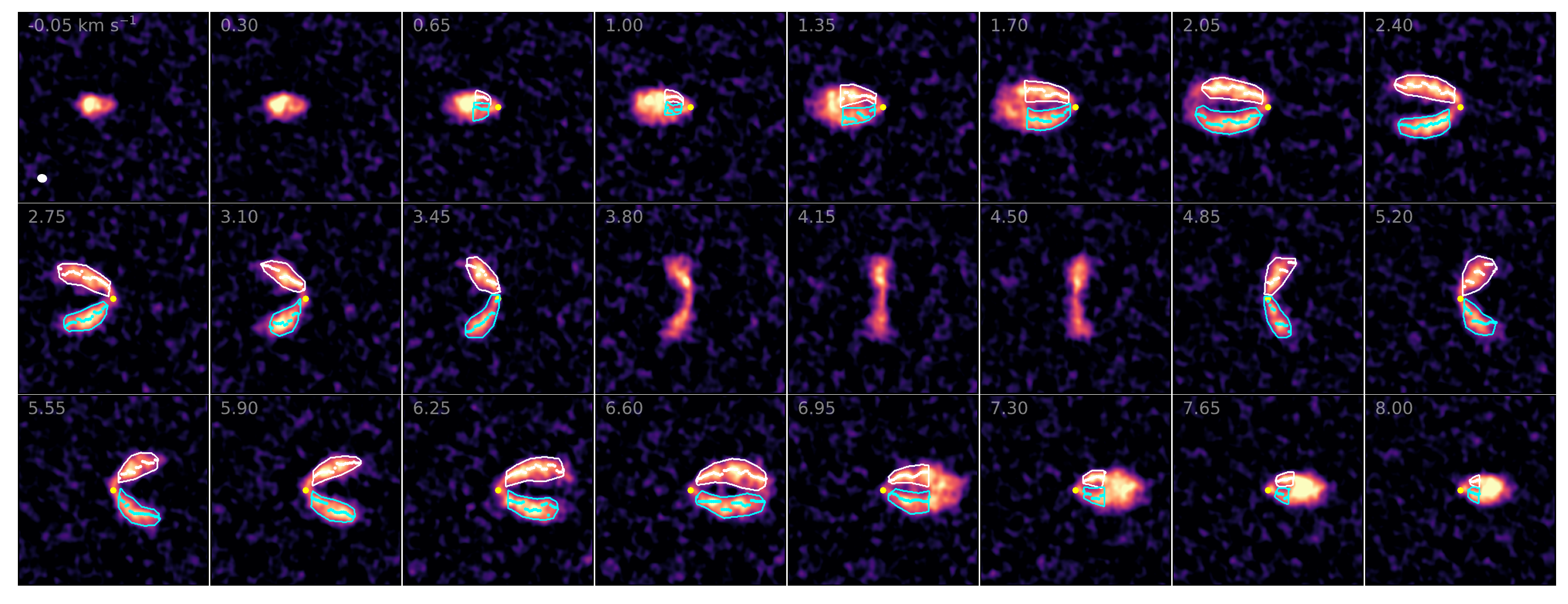}
    \caption{$^{12}$CO channel maps of HD~142666.}
    \label{fig:HD142666_channel_maps}
\end{figure*}

\begin{figure*}
    \centering
    \includegraphics[width=\textwidth]{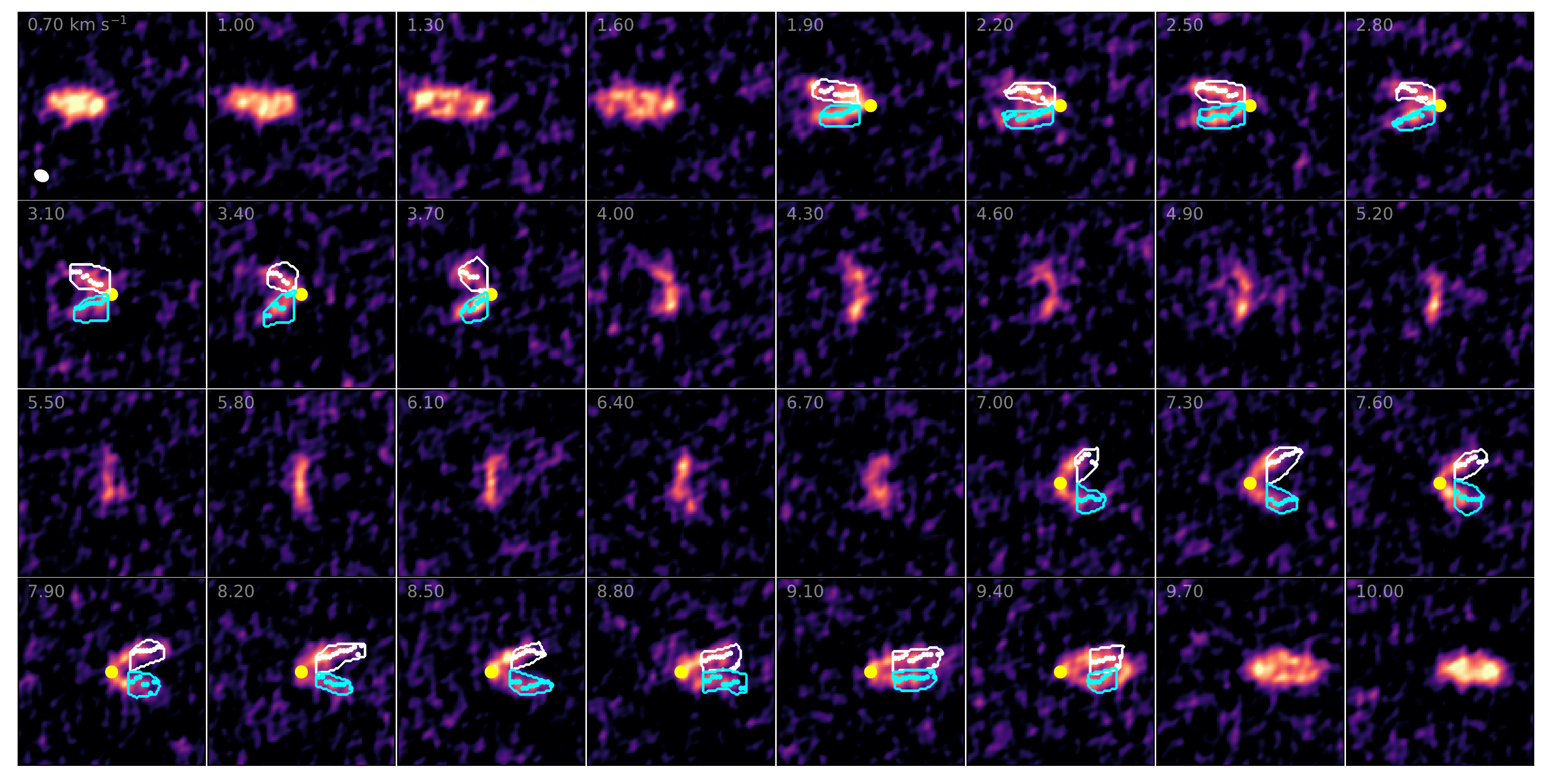}
    \caption{$^{12}$CO channel maps of AK~Sco.}
    \label{fig:AKSco_channel_maps}
\end{figure*}

Figures \ref{fig:HD100546_channel_maps} to \ref{fig:AKSco_channel_maps} show the $^{12}$CO $J=2-1$ channel maps used in this work. The white and blue lines show the outlines of the hand-drawn masks of the far and near side respectively. The white and blue scatter show the corresponding extracted points. The masks have been made by carefully identifying the emitting regions visually. We do not use the central channels and the first and last channels because a horizontal extent and a separation between the far and nearside is necessary to extract the heights. The yellow circle denotes the position of the star. To make the fainter parts better visible, a power-law normalization is used. On each panel, the channel velocity in km~s$^{-1}$ is indicated in the top left corner. The beam size is shown in the bottom left corner of the first panel.

\section{Velocity maps}
\label{app:velocity_maps}
\begin{figure*}
    \centering
    \includegraphics[width=\textwidth]{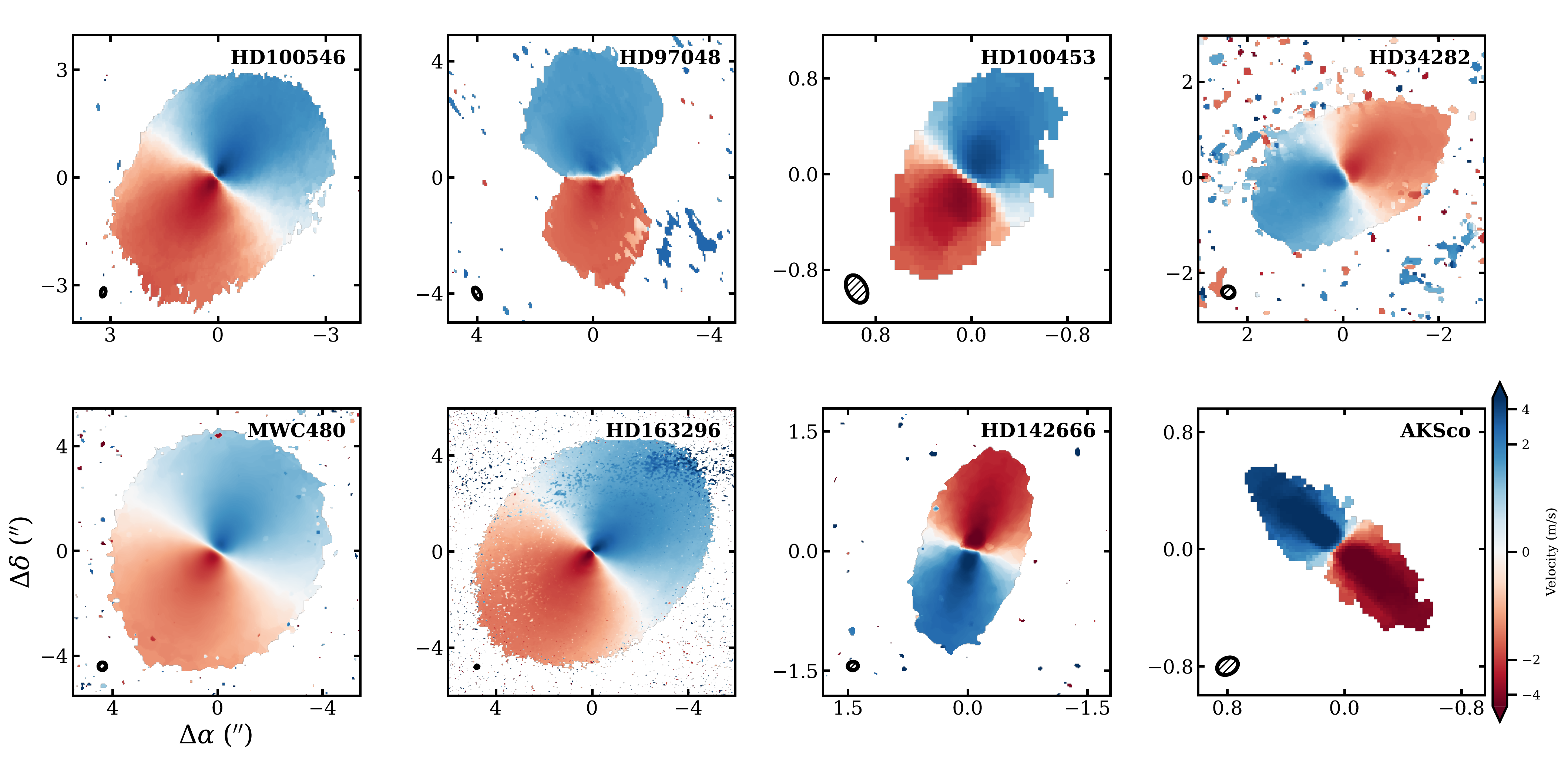}
    \caption{Moment one (intensity weighted average velocity) maps of the eight disks. The top and bottom rows show the group I and II disks respectively. The color bar is centered on their respective system velocities. Each map is clipped at an SNR of 3. The size of the beam is shown in the bottom right corner.}
    \label{fig:mom1}
\end{figure*}

Fig. \ref{fig:mom1} shows the moment 1 maps of the disks, made with \texttt{bettermoments} \citep{Teague2018}, clipped at an SNR of 3. In these figures, the alignment of the red and blueshifted sides provides an additional indication of how vertically extended the disk is. At high enough inclinations, the largest velocities at a particular separation curve with the vertical height of the disk. Hence, if the red and blue shifted sides are aligned in a straight line opposite to each other, the disk is flat, as seen for the HD~142666 and AK~Sco disks. On the contrary, a v-shape indicates a vertically extended disk, as seen in HD~34282 and HD~163296.

\end{appendix}

% \begin{figure}[t]
% \centering
% \begin{subfigure}{.5\textwidth}
%   \centering
%   \includegraphics[width=.3\linewidth]{}
%   \caption{}
%   \label{fig:}
% \end{subfigure}%
% \begin{subfigure}{.5\textwidth}
%   \centering
%   \includegraphics[width=.3\linewidth]{}
%   \caption{}
%   \label{fig:}
% \end{subfigure}
% \caption{\textbf{a)}  \textbf{b)} }
% \label{fig:}
% \end{figure}

\end{document}